\documentstyle[aas2pp4,epsf]{article}

\begin{document}

\title{Inward Propagation of Nuclear-Burning Shells in 
Merging C-O and He White Dwarfs}

\author{ Hideyuki Saio}
\affil{Astronomical Institute, School of Science, 
Tohoku University, Sendai 980, Japan ; saio@astr.tohoku.ac.jp}

\and
\author{Ken'ichi Nomoto}
\affil{Department of Astronomy \& Research Center for the Early Universe, 
School of Science, University of Tokyo,
Bunkyo-ku, 113, Japan, 
and Institute for Theoretical Physics, the University of 
California at Santa Barbara, CA; nomoto@astron.s.u-tokyo.ac.jp}

\begin{abstract}
We have investigated the consequences of merging double white dwarf systems 
by calculating
evolutionary models of accreting white dwarfs.
We have considered two cases; a massive C-O white dwarf of 
$\sim 1M_\odot$ accreting C-O mixture, and
a low mass white dwarf with an initial mass of 
$0.4M_\odot$ accreting matter composed mostly of helium. 

The accretion rate of the C-O white dwarf is assumed to be 
$1\times10^{-5}M_\odot$y$^{-1}$.
After carbon burning is ignited at $M_r\sim1.04M_\odot$,
the flame propagates inward due to heat conduction.
By inserting enough grid points to resolve the structure of the flame,
we have obtained almost steady burning in most phase of evolution, 
but we have found a new phenomenon
that the strength of flame sometimes oscillates due to a {\it thermal
instability} in early phase of the evolution.
In calculating evolutionary models,
we have occasionally employed a steady-burning approximation,
in which the propagation speed of flame is given a priori.
We have considered two extreme cases for the interior abundance of the massive
white dwarf: $X_{\rm C}=0.5$ and $X_{\rm C}=0.2$.
For $X_{\rm C}=0.5$, the flame becomes very weak at $M_r\sim0.4M_\odot$
and the inward propagation stall there. 
But a few thousand years later, the flame is reactivated due to contraction
and propagates to the center.
For $X_{\rm C}=0.2$, the flame propagates smoothly and reaches the center
in $\sim 1000$ years.
In both cases the C-O mixture has been burned into an O-Ne-Mg mixture
without causing an explosive phenomenon.

For helium accreting low mass white dwarfs, we have considered 
accretion rates of
$1\times10^{-7}$ and $1\times10^{-6}M_\odot$y$^{-1}$.
After a fraction of $M_\odot$ is accreted to the white dwarf,
helium is ignited in the outer part and a shell flash occurs. 
Such a shell flash diminishes when about 10\% of helium in the convective
shell is burned into carbon and oxygen.
The next shell flash occurs at a shell interior to the 
previously flashed shell.
After less than 30 shell flashes, the helium ignition occurs at the 
center and steady burning begins.
Thus, the merging produces a helium star which burns helium at the
center. 
The mechanism of the propagation of the burning shell in this case is
compressional heating during inter-pulse phases, in contrast to
the case of carbon-burning flame where the conduction drives the 
inward propagation of the flame.
We infer that such a low mass double white dwarf system could be 
a progenitor of the AM CVn stars.

\end{abstract}


\section{Introduction}

It is expected theoretically that
a fraction of close binary systems become double white dwarf systems
(Iben \& Tutukov\markcite{it84} 1984; Webbink\markcite{wb84} 1984).
Observationary, 
several double white dwarf systems have been found by
Nather, Robinson, \& Stover\markcite{nrs81} (1981),
Saffer, Liebert, \& Olszewski\markcite{slo88} (1988),
Bragaglia, et al.\markcite{bgrd90} (1990),
Marsh, Dhillon, \& Duck\markcite{mdd95} (1995), 
and Marsh\markcite{marsh95} (1995).
Although low mass white dwarfs ($M<0.45M_\odot$) are often found in 
binary systems (Marsh et al. 1995), no massive double white dwarf system
with a short period (less than several hours) has been found despite
several searches by e.g., Robinson \& Shafter\markcite{rs87} (1987),
Foss, Wade \& Green\markcite{fwg91} (1991), and 
Bragaglia et al.\markcite{bgrd90} (1990).
Bragaglia\markcite{bragag95} (1995) discussed the present status of the
search and possible explanations for the null result.
In this paper, we assume that there exist some massive double white
dwarf systems as well as low mass systems.

Double white dwarf systems having short enough periods are expected to
merge within the Hubble time by the effect of gravitational
wave radiation.
Since the radius of a white dwarf is proportional to $M^{-1/3}$,
once the less massive component fills its critical Roche lobe,
a dynamical mass transfer occurs if the secondary to primary
mass ratio is sufficiently large (Cameron \& Iben \markcite{ci86} 1986;
Benz et al.\markcite{benz90} 1990). 
The consequence of the rapid mass transfer seems to be the formation
of a hot thick disk around the primary white dwarf
(e.g., Hachisu, Eriguchi, \& Nomoto \markcite{hen86} 1986; 
Mochkovich \& Livio \markcite{ml90} 1990). 
At least a part of the disk matter eventually falls on the white dwarf
to complete the merging process.
Iben \& Tutukov (1984) and Webbink (1984) argued that  if the total mass
of the two white dwarfs is larger than Chandrasekhar's limiting mass,
the merging would lead to a Type Ia supernova.
Whether it occurs or not, however, depends on the accretion rate onto
the white dwarf, which, in turn, depends on the timescale on which the
angular momentum distribution changes significantly in the thick disk.
As discussed by Mochkovich \& Livio (1990), the timescale can be as short 
as a few days if the thick disk is turbulent.
In such a case, the accretion onto the white dwarf is limited by the
Eddington limit ($\dot M\sim 2\times10^{-5}M_\odot$y$^{-1}$ for 
$M\sim1M_\odot$) due to the effect of radiation pressure. 
For such a high accretion rate, Nomoto \& Iben (1985) 
have shown that carbon-burning is ignited in an outer shell of
the white dwarf due to the effect of compressional heating.
Saio \& Nomoto\markcite{sn85} (1985) and Woosley \& Weaver\markcite{ww86}
(1986) have shown
that carbon-burning ignited near the surface propagates to the center
to burn most of carbon without causing an explosive phenomenon.
In these calculations  the propagation of carbon-burning was simulated
as successive phases of a shell-flash and an inter shell-flash, because
the burning region (flame) was not spatially resolved.

To overcome the poor resolution in stellar evolution models, 
Timmes, Woosley, \& Taam\markcite{timmes94} (1994) calculated the
propagation of flame in initially homogeneous medium with fixed boundary
temperatures. 
They resolved the structure of flame by using an adaptive mesh point method.
They obtained the flame speed as a function of the 
hot side boundary temperature and the cooler side density.
Combining the result with the assumption of balanced power 
between the integrated
nuclear energy generation rate  and neutrino emission rate, they
have concluded that the flame will reach the center if the mass
of the white dwarf is greater than $\sim0.8M_\odot$.
On the other hand, Kawai, Saio, \& Nomoto\markcite{ksn87} (1987)
have shown that carbon burning can ignite in the outer 
part of an rapidly mass-accreting white dwarf 
if it is more massive than $1.07M_\odot$. 
These results indicate that once carbon-burning is ignited the flame
propagates throughout the interior of the white dwarf to convert
the C-O mixture into an O-Ne-Mg mixture.
Therefore, if the merging of a double white dwarf system leads to an
accretion rapid enough to cause an off-center carbon ignition
in the primary white dwarf 
($\dot M\gtrsim 2.7\times10^{-6}M_\odot$y$^{-1}$; Kawai et al. 1987),
no Type Ia supernova is expected.
Since it is very important to know what is the final product of the
merging of double white dwarfs, we have re-investigated the flame propagation
in a white dwarf by inserting enough grid points to resolve the flame 
structure.

In a helium white dwarf accreting helium at a rate faster than
$2\times10^{-8}M_\odot$y$^{-1}$ helium burning is ignited off-center
(Nomoto \& Sugimoto\markcite{ns77} 1977). 
When we made computations for the massive C-O white dwarf
before, we also
calculated helium white dwarf models. 
We found that  helium burning shell
propagates inward to the center by repeating mild flashes,
and that the white dwarf is converted to a helium-main-sequence star.
Some of the results have been presented in Nomoto \& Hashimoto
\markcite{nh86}\markcite{nh87}
(1986, 1987). Iben\markcite{iben90} (1990) investigated similar cases and
claimed that the products of merging low mass white dwarfs are 
sdB and sdO stars (see, e.g., Heber\markcite{heber91} (1991) for a review of
sdB and sdO stars).
In our present paper we also presents results of additional computations for 
helium accreting low mass helium white dwarfs.

\section{C-O white dwarf}

\subsection{Structure of flame}

\begin{figure}
\epsscale{0.8}
\plotone{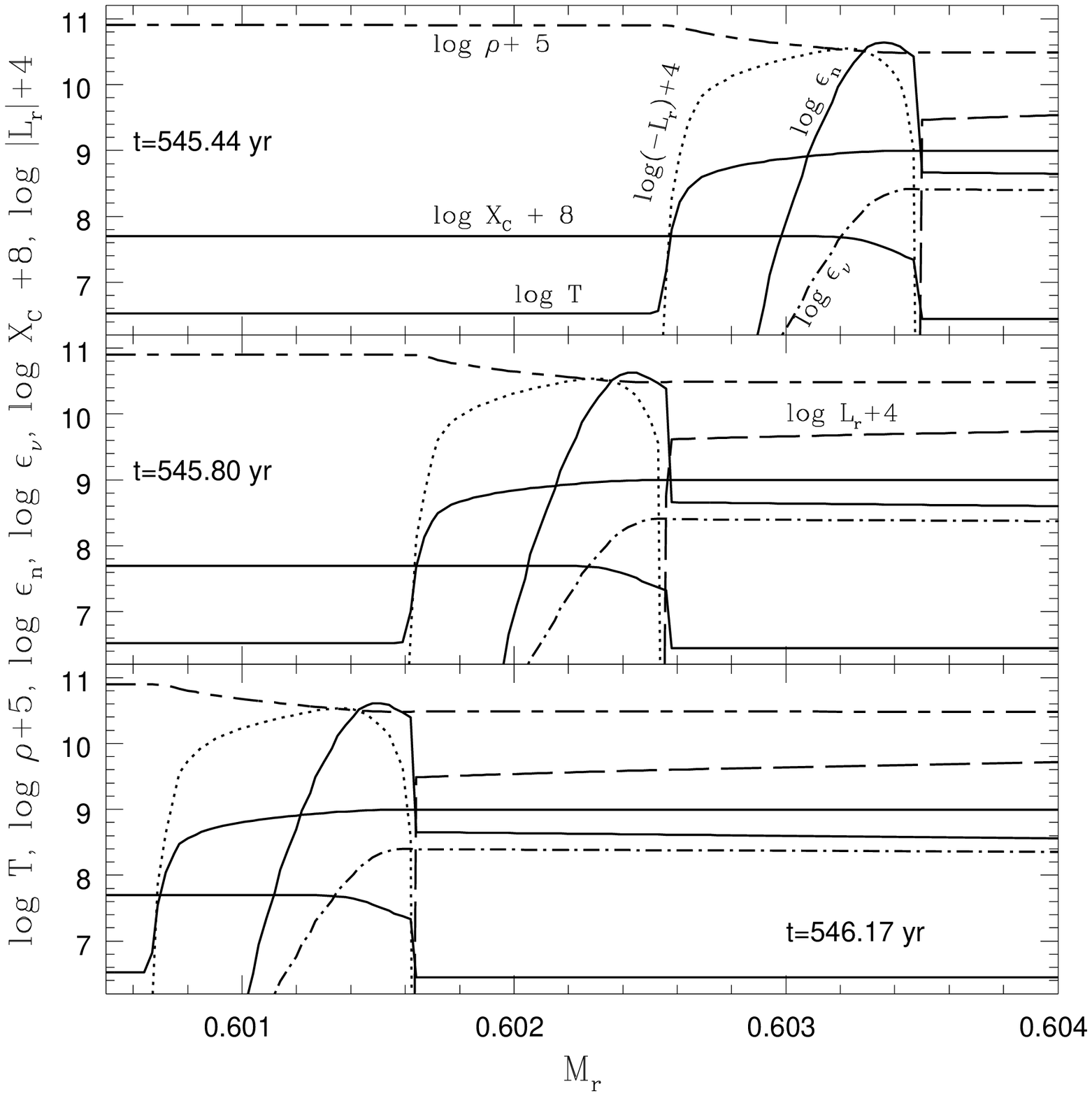}
\caption {Runs of density $\rho$, temperature $T$, carbon abundance
$X_C$, nuclear energy generation rate per unit mass $\epsilon_{\rm n}$,
neutrino emission rate per unit mass $\epsilon_\nu$, and
$L_r$ around the carbon-burning flame
at selected stages of the flame propagation.
To display the variation of $L_r$, the dotted lines are used when $L_r<0$ and
the long dashed-lines when $L_r>0$.
}   
\label{f1}
\end{figure}

 In our previous calculation (Saio \& Nomoto\markcite{sn85} 1985), the 
inward propagation of the carbon-burning shell appeared as a series of
successive shell flashes.
This is an artifact brought by insufficient spatial resolution.
The actual evolution should be a nearly steady progression of the 
burning front or flame (Timmes et al.\markcite{timmes94} 1994).
In this paper, we present the result of our re-computations
for this problem done by inserting enough grid
points to resolve the carbon burning flame.
Input physics and the initial model are the same as in Saio \& Nomoto
\markcite{sn85} (1985).
Figure \ref{f1} shows the flame structure at selected phases 
when the flame is located at $M_r\simeq 0.6 M_\odot$.
The flame, whose center is defined by the peak of nuclear energy generation
rate per unit mass $\epsilon_{\rm n}$,  
moves inward keeping the neighboring structure similar.
Just interior to the flame there is a
thermal front having an extremely steep temperature gradient.
The front moves inward due to the conductive energy flow.
Matter experiences a large heating by the passage of the front. 
The density decreases appreciably between the thermal front 
and the flame center.

\placefigure{f1}

Since the carbon abundance decreases outward in the flame,
the maximum temperature occurs at a zone exterior to the flame center.  
The position of the maximum temperature is recognized in Figure \ref{f1}
as the point where the direction of energy flow changes (i.e., 
the sign of $L_r$ changes; dotted lines are used if $L_r<0$ and long
dashed lines if $L_r>0$ in Fig. \ref{f1}).
Figure \ref{f1} indicates that most of the energy produced in the
flame flows inward and is used to move the thermal front inward.

Behind the flame (i.e., exterior to the flame) a shell convection zone
develops.
The bottom of the convection zone is recognized in Figure \ref{f1}
as sudden decreases of $X_{\rm C}$ and $\epsilon_{\rm n}$.
The convection zone at these stages extends to $M_r\simeq 0.73M_\odot$.
A considerable amount of energy is generated from carbon-burning
in the convection zone.
For the model in the middle panel of Fig. \ref{f1}, 
$L_{\rm n}=7.64\times10^6L_\odot$, 
in which $4.72\times 10^6L_\odot$ is produced in the flame, and the remaining
part is produced in the convective zone,
and $L_\nu=3.06\times 10^6L_\odot$.
Around the outer boundary of the convection zone the outward energy
flow is reduced to $L_r\simeq 1.8\times10^4L_\odot$ due to the
neutrino emission in the convection zone.
Roughly speaking, 40\% of the energy generated by carbon burning is
lost by neutrino emission and the remaining energy is conducted inward
to heat matter between the flame and the thermal front.
Nomoto \markcite{nomoto84}\markcite{nomoto87}(1984,1987) and
Garc\'ia-Berro, Ritossa, \& Iben \markcite{gri97} (1997) have obtained  similar
results for a carbon-burning flame in the asymptotic giant branch phase of
a $9M_\odot$ model.

\subsection{Propagation of flame}
Keeping grid spacing fine enough to resolve the thermal front and the
flame demands very small time-steps to follow the flame propagation.
To alleviate the computational effort we used a steady burning approximation 
occasionally in evolutionary computations. 
In the steady burning approximation, we do not resolve the structure around
the flame; i.e., the thermal front and the flame is represented
by a grid point.
The total nuclear energy generation rate
$L_{\rm n}$ is calculated by 
\begin{equation}
L_{\rm n}=X_{\rm C}Q|d{m}_{\rm f}/dt|,
\end{equation}
where $Q$ is the energy yield from carbon burning, for which 
we have adopted $Q=4.4\times10^{17}$ergs s$^{-1}$g$^{-1}$, and
$m_{\rm f}$ is the mass coordinate of the flame center. 
The local energy generation rate per unit mass $\epsilon_{\rm n}$ was set to be
zero except at the grid point corresponding to the flame; i.e.,
$\epsilon_{{\rm n}j}=0$ if $j\neq J$ and 
$\epsilon_{{\rm n}J}=L_n/(M_{r,J+1}-m_{\rm f})$,
where $J$th grid point corresponds to the flame and $M_{r,j}$ 
represents the mass coordinate at the $j$th grid point.
The value of $\dot{m}_{\rm f}$ for the steady burning calculations was 
obtained from a few thousand preceding models 
obtained by usual evolution calculations with fine grid spacings.

In our present investigation, evolutionary calculations were started with
a model from Saio \& Nomoto\markcite{sn85} (1985).
The model has a carbon-burning flame at $m_{\rm f}\sim0.85M_\odot$.
Interior to the flame, carbon and oxygen abundances are 
($X_{\rm C}$,$X_{\rm O}$)=(0.5,0.5). 
To see the effect of the carbon abundance in the interior, we 
have also computed a set of evolutionary models 
starting with the initial model
in which the abundance interior to the flame had been artificially 
converted to ($X_{\rm C}$,$X_{\rm O}$)=(0.2,0.8). 
In both cases, the accretion rate was kept constant 
at $10^{-5}M_\odot$y$^{-1}$.

\begin{figure}
\epsscale{0.8}
\plotone{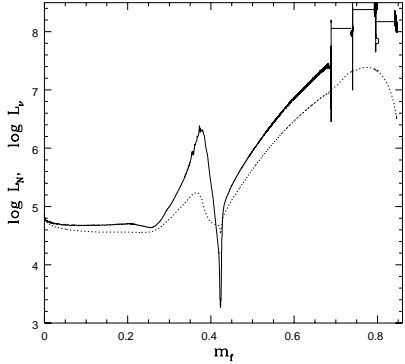}
\caption{Total nuclear energy generation rate $L_{\rm n}$ (solid line)
and neutrino energy loss rate $L_\nu$ (dotted line) are shown as functions
of $m_{\rm f}$ (the mass coordinate at the carbon burning flame) for 
$X_C=0.5$. The phases with constant $L_{\rm n}$ correspond to the phases
during which the steady burning approximation was applied.
Large fluctuations in $L_{\rm n}$ at the ends of such phases
are artifacts caused by inserting many grid points around the flame.
}
   \label{f2}
\end{figure}

\begin{figure}
\epsscale{0.8}
\plotone{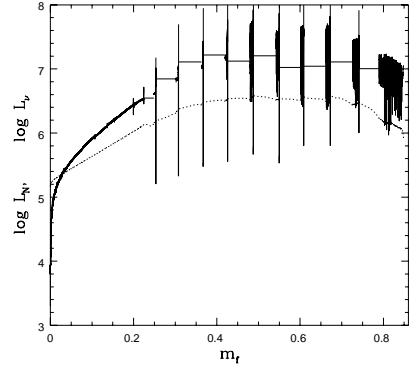}
\caption{The same as Fig. 2
but for $X_{\rm C}=0.2$
}
   \label{f3}
\end{figure}

Figures \ref{f2} and \ref{f3} show evolutionary changes
in the nuclear energy generation rate $L_{\rm n}$ (solid line)
and the neutrino energy loss rate $L_\nu$ (broken line)
as a function of the mass coordinate of the carbon-burning
flame $m_{\rm f}$ for $X_C=0.5$ and $0.2$, respectively.
In these figures, the phases with constant $L_{\rm n}$ 
correspond to the evolution phases during which the steady burning 
approximation was applied.
As can be seen from these figures, one set of calculations with the
steady burning approximation was stopped 
when $m_{\rm f}$ changed by $\sim 0.05M_\odot$.
Large fluctuations in $L_{\rm n}$ at the transitions from the steady burning
calculation to the normal evolutionary calculation 
are artificial phenomena caused by inserting
grid points just interior to the flame.

\placefigure{f2}
\placefigure{f3}

In most phase of the progression of the carbon-burning flame the nuclear
energy generation rate $L_{\rm n}$ 
exceeds the neutrino energy loss rate $L_\nu$.
As discussed above, the difference is approximately 
the amount of energy conducted inward.
This energy balance is similar to that found by Garc\'ia-Berro et al.
(1997) but somewhat different from the balanced power assumption
employed by Timmes et al.\markcite{timmes94} (1994).
For $X_{\rm C}=0.5$, the inward propagation of the flame stalls at
$M_r\sim 0.4M_\odot$, which appears as a sharp dip of $L_{\rm n}$ in 
Fig. \ref{f2}.
We discuss this phenomenon more below.
In both cases the flame reaches to the center and exhausts carbon
interior to the first ignition shell, which confirms our previous 
conclusion (Saio \& Nomoto 1985).

\begin{figure}
\epsscale{0.8}
\plotone{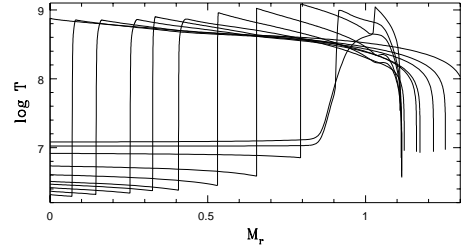}
\caption{Variation of the temperature distribution as the carbon-burning
flame propagates inward.}
   \label{f4}
\end{figure}

\begin{figure}
\epsscale{0.8}
\plotone{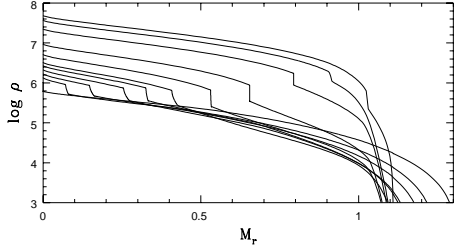}
\caption{Variation of the density distribution  as the 
carbon-burning flame propagates inward.}
   \label{f5}
\end{figure}

Figures \ref{f4} and \ref{f5} show the time variation of the distributions 
of temperature and of density, respectively,
during the flame propagation for $X_{\rm C}=0.5$.
The flame temperature is maximum when the burning front is located at
$M_r\simeq0.8M_\odot$, i.e., when $m_{\rm f}\simeq0.8M_\odot$. 
Then it decreases gradually as the flame propagates toward the center.
Since pressure is continuous across the flame, density decreases
as matter gets into the flame.
The density just interior to the flame is maximum when $m_{\rm f}\simeq
0.9M_\odot$.
It decreases as the flame propagates inward until it becomes minimum when
$m_{\rm f}\simeq0.5M_\odot$, and then increases slowly as the flame approaches 
the center.
These variations depend on the flame speed, bulk expansion rate of the
white dwarf, density gradient, the degree of electron degeneracy, etc.

\placefigure{f4} 
\placefigure{f5}

The central temperature and density decrease as the 
flame propagates.
This variation is caused by the
net gain of energy because the nuclear energy generation rate exceeds
the neutrino loss rate in most phases of the flame propagation.
(When the neutrino loss rate exceeds the nuclear energy generation rate,
both the central temperature and density increase [see below].)
The decrease of the central temperature seen in Fig. \ref{f4}
is larger than the case shown in Saio \& Nomoto\markcite{sn85} (1985).
This can be explained as follows: In Saio \& Nomoto\markcite{sn85} (1985),
the spatial resolution was not enough so that the flame propagation 
appeared as a series of shell-flash and inter shell-flash phases.
During an inter shell-flash phase, the neutrino loss rate exceeded the
nuclear energy generation rate.
Therefore, the net gain of energy in the present calculation is 
higher than the case in Saio \& Nomoto (1985).

\begin{figure}
\epsscale{0.8}
\plotone{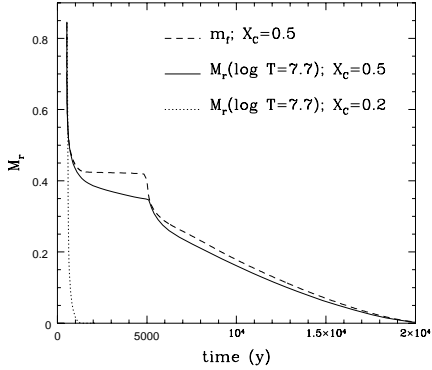}
\caption{Changes in the position of flame (dashed line) and
thermal front (solid line) for $X_{\rm C}=0.5$.
The dotted line indicates the thermal front for $X_{\rm C}=0.2$, for which
the flame is always very close to the thermal front.} 
\label{f6}
\end{figure}

The time variation of $m_{\rm f}$ for $X_{\rm C}=0.5$ 
is shown in Figure \ref{f6} by the dashed line.
The solid line and the dotted line in this figure show the position of 
thermal front for $X_{\rm C}=0.5$ and 0.2, respectively. 
The shell having $\log T=7.7$ well represents the position of the thermal 
front, which is located just interior to the flame (see Fig. \ref{f1}).
The propagation speed of the flame for $X_{\rm C}=0.5$ 
decreases significantly and the flame stalls for a while 
at $M_r\sim 0.4M_\odot$. 
This phase
corresponds to a deep dip in $L_{\rm n}$ seen in Fig. \ref{f2}. 
Even when the flame propagation stops, the thermal front gradually moves 
inward but the steepness of the temperature gradient decreases
(Fig. \ref{f7} below).
After the flame have stalled there for $\sim 4000$ years, 
it becomes active again
and propagates toward the center at a speed considerably slower than before.
After the re-activation, the carbon abundance in the convective 
shell is $\sim0.1$, which is 
considerably higher than the case before the flame stalled.
After the flame reached the center, the remaining carbon has been burnt  
quietly in a few thousand years.

\placefigure{f6}

The flame for $X_{\rm C}=0.2$, on the other hand, reaches the center smoothly
in $\sim 1000$ years and the thermal front keeps very close to the flame
during the propagation.
Since the flame remains strong throughout the propagation in this case,
the carbon abundance in the convective zone never exceed $0.005$.
Therefore, when the flame reached the center, most of carbon in the region
interior to the first ignition shell has been exhausted.

\begin{figure}
\epsscale{0.8}
\plotone{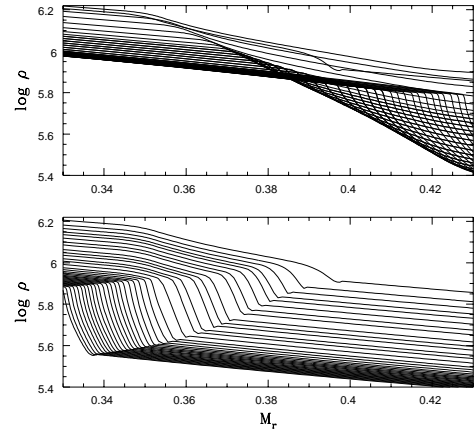}
\caption{Evolution of density distribution before (upper panel) and
after (lower panel) the re-activation of flame.}
\label{f8}
\end{figure}

Figures \ref{f7} and \ref{f8} show variations of 
temperature and density distributions, respectively, in the phase during
which the propagation of the flame almost stopped and was re-activated later 
for $X_{\rm C}=0.5$.
Upper and lower panels of these figures show evolution 
before and after the re-activation of flame, respectively.
The direction of evolution is recognized by the inward propagation
of the thermal front. 
The last lines in the upper panels are the same phase as the first lines
in the lower panels.
Before the re-activation, as the thermal front propagates inward,
the temperature gradient at the front 
becomes less steep and the density jump gets smaller and less apparent.
During this phase, the flame becomes weak and hardly moves at
$M_r\simeq0.42M_\odot$ (Fig. \ref{f6}).
Since $L_{\rm n}<L_\nu$ in the phases shown in the upper panels of 
these figures, the density interior to the front increases with time.
The burning becomes active again at $M_r\sim0.4M_\odot$, when the thermal front
is located around $M_r\sim0.35M_\odot$.
A new thermal front backed  by the activated flame appears outside of the old
front and propagates inward to engulf the old one 
at $M_r\simeq0.345M_\odot$.

\placefigure{f7}
\placefigure{f8}

The time spent for the flame to reach the center is shorter
for $X_{\rm C}=0.2$ than the case of $X_{\rm C}=0.5$.
Moreover, for $X_{\rm C}=0.5$ the flame in the present calculation
takes longer to reach the center than in our
previous calculation (Saio \& Nomoto\markcite{sn85} 1985), in which
it took about 5000 years for the flame to propagate through the interior.
Such differences can be explained by the fact that the net-energy
gain ($L_{\rm n}-L_\nu$) in the present $X_{\rm C}=0.5$ case is larger than
the $X_{\rm C}=0.2$ case or our previous calculation.
A larger energy-gain makes gentle the temperature increase in the inner
part of the flame, because
a larger energy-gain decreases  more the density at the ignition temperature,
thus making the degree of degeneracy there lower.
A lower temperature gradient makes the propagation speed lower.

\subsection{Discussion}
\begin{figure}
\epsscale{0.8}
\plotone{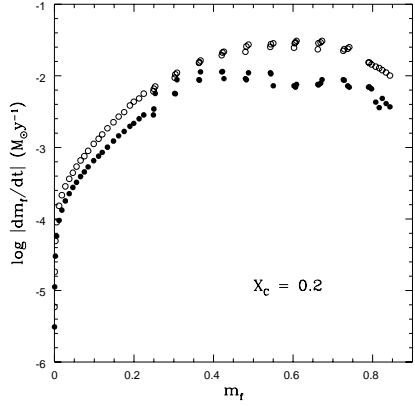}
\caption{Propagation speed $|dm_{\rm f}/dt|$ ($M_\odot$y$^{-1}$) 
versus position $m_{\rm f}$ of the flame for $X_{\rm C} = 0.2$.
Filled circles represent propagation speed evaluated from the propagation
of the flame in evolutionary models, while open circles represent the 
values obtained by interpolating
Timmes et al.'s  (1994) table for the 
density and temperature at the flame.}
   \label{f9} 
\end{figure}

Timmes et al.\markcite{timmes94} (1994) evaluated propagation speeds of 
carbon-burning flames in a isobaric region which has initially homogeneous
temperature and density.
They obtained flame speed in such systems as a function of the boundary
temperature of the hotter side and the density of the cooler side 
(initial density) of the flame.
The boundary temperature in the 
cooler side (i.e., initial temperature) was fixed at $5\times10^8$K, 
which is much higher than in our models.
Figure \ref{f9} shows the propagation speed $|d m_{\rm f}/dt|$ of the flame
derived from the actual propagation in our evolutionary models for 
$X_{\rm C}=0.2$ 
(filled circles) and that taken 
from the table of propagation
speed for $X_C=0.2$ in Timmes et al.\markcite{timmes94} (1994)
for the density and temperature at the flame in 
our models (open circles).
This figure indicates that the propagation speed based on Timmes et al.'s
result tends to be higher than that in our evolutionary models.
One reason for this difference may be ascribed to the fact that 
the cooler side 
temperature of our models is much lower than that in Timmes et al.'s analyses.
We note here that evolution calculations using the steady state 
approximation with
the propagation speed of Timmes et al. failed before the flame reaches the
center because the density became too low.
The rapid density decrease occurred because the energy production rate was 
too high, which indicates that their propagation speed seems too high
for our white dwarf models. 

\placefigure{f9}

\begin{figure}
\epsscale{0.8}
\plotone{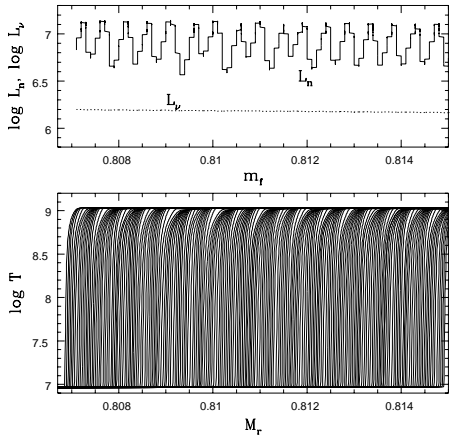}
\caption{The upper panel shows $L_{\rm n}$ and $L_\nu$ as functions of 
the mass coordinate of the flame for $X_{\rm C} = 0.2$. 
Time runs from right to left.
The lower panel shows temperature distribution near flames at 
various evolutionary stages.}
   \label{f10}
\end{figure}

\begin{figure}
\epsscale{0.8}
\plotone{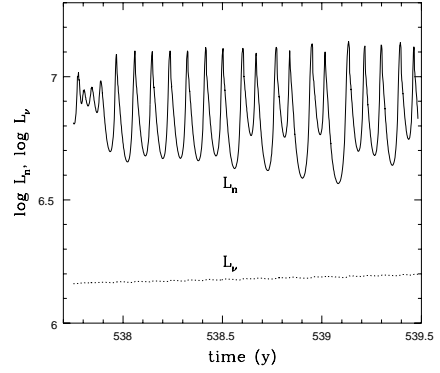}
\caption{The time variations of $L_{\rm n}$ and $L_\nu$ during the
 evolution phase shown in Fig. 10.}
 \label{f11}
\end{figure}

We have found a new phenomenon that the propagation of the
flame tends to be oscillatory when the peak temperature is relatively high.
An example is shown in Figs. \ref{f10} and \ref{f11}. 
The lower panel of Fig. \ref{f10} shows runs of temperature
in the interior for every 20 evolutionary models, 
while the upper panel shows nuclear energy generation rate $L_{\rm n}$ 
and neutrino emission rate $L_\nu$ 
as functions of $m_{\rm f}$
(time advances from right to left of the diagram).
Figure \ref{f11} shows $L_{\rm n}$ and $L_\nu$ 
as functions of time.
As seen in Fig \ref{f10}, the temperature gradient between
the ignition point ($\log T\simeq 8.8$) and the position of maximum
temperature becomes gentle occasionally, and
the energy generation rate dips.
The time variation of $L_n$ is characterized by semi-regular oscillations
with sharp peaks and relatively gentle minimums.
The typical `period' is an order of $0.1$y. 
In contrast to the variation of $L_n$, $L_\nu$ hardly changes.

\placefigure{f10} 
\placefigure{f11}

Such oscillatory behavior of the flame tends to occur 
when the thermal front is resolved by finer grid spacings.
Since finer grid spacings make the maximum temperature steepness of
the thermal front larger, we infer that a steep temperature 
gradient plays an important role in the instability.
We may explain the cause of 
the fluctuation as follows:
When the temperature gradient becomes too steep,
the inward energy flow exceeds the energy generation by 
nuclear burning to decrease the flame temperature and hence $L_n$. 
This decreases the steepness of temperature distribution,
which in turn slows the inward energy flow.
Then, the flame temperature and $L_n$ increase to recover the previous
levels.

It is known that a combustion flame front in a solid medium (in which the
diffusion rate of the fuel is much slower than the rate of 
thermal conduction) is thermally unstable if the temperature 
sensitivity of the energy generation rate is sufficiently high 
(see e.g., Zeldovich et al \markcite{zblm85} 1985).
The high temperature sensitivity of energy generation rate makes the flame
thin with a steep temperature gradient.
We believe that the instability is the same as that of 
the carbon-burning flame in our accreting white dwarf because
both of them occur under the condition of a steep temperature gradient
and a high temperature sensitivity of energy generation rate.

\section{Heium white dwarf}

The starting model adopted for a helium-accreting white dwarf is
a chemically homogeneous (Y=0.98, Z=0.02) white dwarf of $0.4M_\odot$.
The white dwarf is assumed to be the massive component of a double
white dwarf system.
When the less massive component fills the critical Roche lobe
(because of the gravitational wave radiation),
the mass transfer is unstable if the mass ratio exceeds 2/3.
In this case the massive component is expected to accrete matter
at a rate of the Eddington limit, which is of the order of 
$10^{-5}M_\odot$yr$^{-1}$.
If the mass ratio is less than 2/3, stable mass transfer occurs at
a rate of e.g., $3\times10^{-6}$yr$^{-1}$ for $M_2=0.2M_\odot$,
and $5\times10^{-8}M_\odot$yr$^{-1}$ for $M_2=0.1M_\odot$, where
$M_2$ indicates the mass of the less massive component.
Iben\markcite{iben90} (1990) investigated the consequences of
rapid merger processes in which mass is accreted at a rate faster
than the critical Eddington limit.
We investigate here cases with mass accretion rates  
$10^{-7}M_\odot$yr$^{-1}$ and $10^{-6}M_\odot$yr$^{-1}$, which
are much smaller than the critical Eddington limit.
 
The chemical composition of accreting matter is set to be 
$(Y,Z)$=(0.98,0.02) assuming that  the less massive component is also
a helium white dwarf.

\subsection{Results}

Accretion causes heating in the envelope of the white dwarf.
When a sufficient mass is accreted, the triple alpha reaction is ignited.
For higher accretion rate the total mass at the helium ignition 
is lower and the mass between the ignition shell and the surface 
is smaller.
The ignition occurs at $M_r\simeq0.44M_\odot$ when the total mass becomes
$M\simeq0.52M_\odot$ for $\dot M=10^{-7}M_\odot$y$^{-1}$; 
these quantities are, respectively, $M_r\simeq0.41M_\odot$ 
and $M\simeq0.45M_\odot$
for $\dot M=10^{-6}M_\odot$y$^{-1}$.
The ignition of the triple alpha reaction leads to 
a shell flash (thermal pulse). 

\begin{figure}
\epsscale{0.8}
\plotone{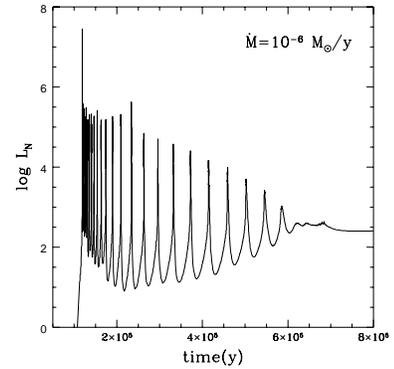}
\caption{The time variation of the integrated nuclear energy generation
rate during the process of inward propagation of the helium burning
shell for  $\dot M=10^{-6}M_\odot$y$^{-1}$.}

   \label{f12}
\end{figure}

After the shell flash is over, another shell flash occurs at a slightly
inner shell than the previous one.
Such shell flashes repeat $\sim20$ to 30 times 
until the ignition shell reaches 
the center of the star.
The number of shell flashes is larger than for the cases investigated by
Iben\markcite{iben90} (1990).
The difference is attributed to the fact that the first ignition shell
in Iben's models ($M_r\sim0.25M_\odot$) is closer to the center 
than our models. 
Figure \ref{f12} shows the time variation of the integrated 
nuclear energy generation rate $L_{\rm n}$ in solar units
for $\dot M=10^{-6}M_\odot$y$^{-1}$.
The strength of a shell flash decreases as the ignition shell
moves inward.
The evolution of repeating mild flashes is similar to the 
evolution which occurs after
the ignition of helium core flash in a red giant star 
(e.g., Mengel \& Sweigart\markcite{ms81} 1981).

\placefigure{f12}

The first shell flash is stronger for the lower accretion rate,
because the density at the ignition shell is higher.
At the peak of the first shell flash $\log L_{\rm n}$ reaches
up to $\sim7.5$ for $\dot M=10^{-6}M_\odot$y$^{-1}$, while
$\sim10.3$ for $\dot M=10^{-7}M_\odot$y$^{-1}$.
The strength of the second and the later shell flashes hardly
depends on the accretion rate. 

\begin{figure}
\epsscale{0.8}
\plotone{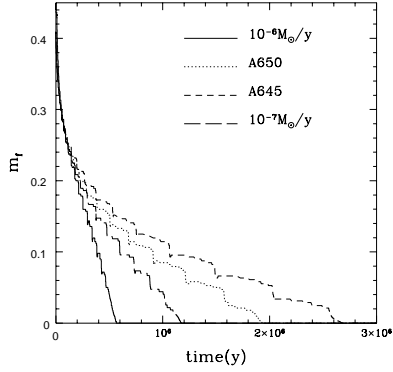}
\caption{The mass coordinate at the helium burning shell versus time after
the ignition.}
   \label{f13}
\end{figure}

Figure \ref{f13} shows the time variation of 
the mass coordinate $m_{\rm f}$ where
the nuclear energy generation rate per unit mass is maximum.
The zero point of time in this figure is set at the first ignition
of helium burning.
The helium burning shell reaches the center in $\sim 5.7\times10^5$y
for $\dot M=10^{-6}M_\odot$y$^{-1}$ and in $\sim1.2\times10^6$y for
$\dot M=10^{-7}M_\odot$y$^{-1}$ after ignition.
The inward move of $m_{\rm f}$ is faster for  
higher accretion rate because the total mass is larger.
The total mass increases to about $1M_\odot$ before the burning
shell reaches the center for $\dot M=10^{-6}M_\odot$y$^{-1}$, while 
$0.63M_\odot$ for $\dot M=10^{-7}M_\odot$y$^{-1}$.
At each shell flashes, a convection zone appears above the burning shell.
One shell flash ends when the helium abundance in the convection
zone decreases to $\sim0.90$.
Therefore, when the burning shell reaches the center, the star
becomes to have a structure similar to that of the 
zero-age helium main-sequence.

\placefigure{f13}

The inward propagation of the burning shell does not stop even if
the accretion stops.
Figure \ref{f13} includes two cases in which accretion 
has stopped before the
burning shell reaches the center. In one case, A6S45, 
the accretion ($\dot M=10^{-6}M_\odot$y$^{-1}$) has stopped
when the total mass became $0.45M_\odot$ (just after the first
shell flash) and in another case, A6S50, the accretion has 
stopped when the total mass became 
$0.5M_\odot$ (when the burning shell was located at $M_r\simeq 0.3M_\odot$).
The time in which helium burning reaches the center for A6S45
is about $2.7\times10^{6}$, which is longer than for A6S50 and is 
similar to Iben's (1990) $0.46M_\odot$ case.
Since the accretion process assumed in Iben (1990) is quite different
from ours, we can conclude that the time is mainly determined by the
total mass and is insensitive to the accretion rate itself.

\begin{figure}
\epsscale{0.8}
\plotone{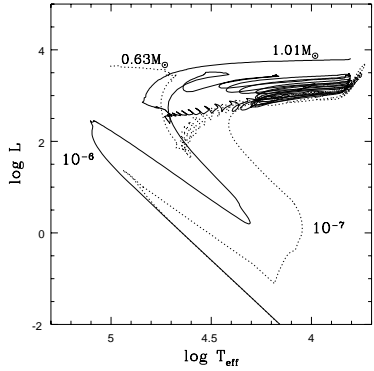}
\caption{The HR diagram showing evolutionary paths of accreting 
white dwarfs with
$\dot M=1.0\times10^{-6}M_\odot$y$^{-1}$ (solid line) and
$\dot M=1.0\times10^{-7}M_\odot$y$^{-1}$ (dotted line).
(The accretion is terminated when the burning shell reaches the center.)
}
   \label{f14}
\end{figure}

Figure \ref{f14} shows the evolutionary paths of our accreting
helium white dwarfs in the HR diagram.
After accretion starts the luminosity increases along the white
dwarf sequence until the triple-alpha reaction is ignited.
Once the ignition occurs, the luminosity decreases at first and
increases later. By the end of the first shell flash,
the star moves into the super-giant region.
From this behavior we can infer that even if the mass transfer is stable
and $\dot M$ is as low as $1\times10^{-7}M_\odot$y$^{-1}$, once
helium burning is ignited, the radius increases to engulf the companion.
As the burning shell moves inward with repeating shell flashes,
the star gradually evolves blueward to the  
helium-main-sequence.
The redward extension of the evolutionary paths during the phase of 
repeating flashes is larger than Iben's (1990) models, because 
in the latter models the mass above the first ignition shell
($>0.1M_\odot$) is larger than in our models 
($\sim 0.04M_\odot$ for $\dot M=1\times10^{-6}$, and
$\sim 0.07M_\odot$ for $\dot M=1\times10^{-7}$).

\placefigure{f14}

When the burning shell reached the center, the accretion rate was
set to be zero. 
Then, the total mass is $0.63M_\odot$ for $\dot M=1\times10^{-7}$
and $1.01M_\odot$ for $\dot M=1\times10^{-6}$.
After the burning shell has reached the center, 
the star spends core helium
burning phase as a helium-main-sequence star which lasts
of order of $10^7$y.
The star evolves off the helium main-sequence when the helium is
exhausted in the core.
The star with $0.63M_\odot$ (dotted line) 
evolves to become brighter and bluer and will evolve along with 
the white dwarf cooling sequence.
The star with $1.01M_\odot$ 
evolves to super-giant region again; the  
behavior is known from the helium star evolution calculations by
e.g., Paczy\'nski\markcite{pac71} (1971) and Saio\markcite{saio95} (1995).

\subsection{Discussion}
\subsubsection{Comparison with the case of carbon-oxygen white dwarfs}

\begin{figure}
\epsscale{0.8}
\plotone{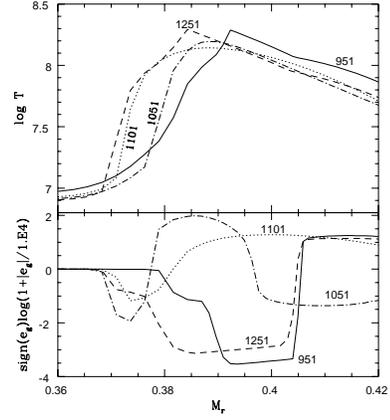}
\caption{Runs of temperature (upper panel) and the rate of gravitational
energy release ($\epsilon_{\rm g}$, lower panel) around the
temperature maximum at some selected stages during one cycle of
a shell flash (from a peak of shell burning to the next one).}
   \label{f15}
\end{figure}

Figure \ref{f15} shows runs of the temperature and the rate of 
gravitational energy 
release $\epsilon_{\rm g}$ around the burning shell at selected stages 
from the peak of a shell flash to the next one.
The number for each curve indicates the corresponding model number.
At the peak of a shell flash a sharp peak appears in the temperature
distribution (\#951 and \#1251), and a zone exterior to the temperature
peak has negative $\epsilon_{\rm g}$, which indicates that 
it is expanded by absorbing heat released by the helium burning.
This figure also shows that during the inter-flash period, the layer
beneath the flashed shell is heated by compression 
(i.e., $\epsilon_{\rm g}> 0$).
The next shell flash occurs at a shell in the heated layer.
Thus, the main driving force of the inward propagation of the burning shell
is the compressional heating during the inter-flash phase.
This differs from the case of the carbon burning flame.
In the latter case, as discussed in the previous section, 
inward conductive heat flow is the main cause for the flame propagation.

\placefigure{f15}

In the case of carbon burning flame the burning shell exhausts carbon
during the inward propagation and most of the released nuclear 
energy is conducted inward or lost by neutrino emission in the
convection zone just above the flame.
In the case of helium burning shell, on the other hand, only a small fraction
of helium is converted to carbon and oxygen during the inward
progression of the burning shell and neutrino emission rate is 
always much smaller than the nuclear energy release.

\subsubsection{Relation with observed double white dwarf systems}
Several double white dwarf systems which consist of low mass 
($\lesssim0.4M_\odot$) white dwarfs have been found 
(Saffer, Liebert, \& Olszewski\markcite{slo88}  1988, 
Bragaglia, Greggio, \& Renzini\markcite{bgr90} 1990, 
Marsh, Dillon, \& Duck\markcite{mdd95} 1995, Marsh\markcite{marsh95} 1995).
The secondary components of these systems seem to have comparable masses.
Since the masses of those white dwarfs are too small to ignite
the triple alpha reaction by the effect of their own gravity,
these are probably helium white dwarfs.
Although all the systems are detached, some of the systems have orbital
periods of a few hours and are expected to become semi-detached systems
in a few billion years (Marsh et al. 1995, Marsh 1995).

When the less massive component of a double white
dwarf system exceeds its critical Roche lobe, mass transfer starts.
If the mass ratio of the secondary to the primary is higher than about 2/3,
mass transfer on the dynamical time scale is expected to occur
(at least in the beginning) to form a common envelope, 
and the primary accretes matter at a rate of the critical Eddington limit.
Even if the mass ratio is smaller and mass transfer is stable,
a common envelope is expected to be formed because
the radius of the primary becomes the supergiant size 
after the triple alpha reaction is ignited in the primary as discussed above
(Fig. \ref{f14}).

We do not know exactly what is the consequence of the common envelope phase.
We infer that matter is peeled off from the secondary.
If the system loses angular momentum effectively enough, 
the separation between the primary and the secondary decreases to merge
(i.e., `spiral-in' occurs).
The system would become a single helium star 
with a mass of $\lesssim0.8M_\odot$, which will eventually become
a DB white dwarf.

If the systemic loss of angular momentum is not large, the
separation can increase as the mass transfer proceeds.
In this case, common envelope phase ends when the radius of the primary
shrinks and the remnant of 
the secondary gets out of the primary envelope.
Considering the fact that there exist recurrent novae, 
in which in each nova explosion
the secondary component is engulfed in the expanding envelope, we infer that
the secondary in the double white dwarf system can survive the common
envelope phase.
In the double white dwarf system, however, the separation is much smaller
and the duration of the common envelope phase is much longer than 
in the case of novae.
Therefore, it is expected that a significant amount of mass is peeled off
from the secondary. If the secondary survives,
the remaining system would have a primary in the helium main sequence
phase with a mass less than $0.8M_\odot$ and a secondary white dwarf
having a very small mass.
After helium is exhausted in the primary, the system becomes
a double white dwarf system with a small mass ratio.
Such a system loses its angular momentum by the gravitational wave
emission and eventually the less massive component fills its critical
Roche lobe to resume mass transfer.
(Because of the small mass ratio, the mass transfer is dynamically stable.)
Such interacting double white dwarf systems can be identified as 
AM CVn type stars, which are hydrogen-deficient cataclysmic variables
composed of double white dwarfs with a small mass ratio ($M_2/M_1 < 0.1$;
see Warner\markcite{warner95} (1995) for a thorough 
review on the AM CVn stars).

\section{Conclusion}

We have investigated the consequences of merging double white dwarf systems 
by calculating evolutionary models for accreting white dwarfs. 
We have considered two cases; a massive C-O white dwarf of 
$\sim M_\odot$ accreting C-O mixture, and
a low mass white dwarf with a initial mass of $0.4M_\odot$.

For the C-O white dwarf an accretion rate of $1\times10^{-5}M_\odot$y$^{-1}$
was assumed.
After the carbon burning is ignited at $M_r\sim1.04M_\odot$,
the flame propagates inward due to heat conduction.
We have considered two extreme cases for the interior abundance of the massive
white dwarf: $X_{\rm C}=0.5$ and $X_{\rm C}=0.2$.
For $X_{\rm C}=0.5$, the flame becomes very weak at $M_r\sim0.4M_\odot$
and the inward propagation stalls there. 
But a few thousand years later, the flame is reactivated due to contraction
and propagates to the center.
For $X_{\rm C}=0.2$, the flame propagates smoothly and reaches the center
in $\sim1000$ years.
In both cases, 
most of the interior of the white dwarf has been converted from
the C-O mixture into an O-Ne-Mg mixture without causing an explosive
phenomenon.
Therefore, if the merging of a double C-O white dwarf system 
leads to an accretion onto the primary white dwarf rapid enough to
ignite carbon off-center,
we cannot expect a Type Ia supernova 
but the formation of a neutron star (Saio \& Nomoto 1985; 
Nomoto \& Kondo 1991).
We have found a new phenomenon that the flame propagation
tends to be oscillatory when the peak temperature is relatively high.

For the helium white dwarf we have considered
accretion rates $10^{-7}M_\odot$y$^{-1}$ and
$10^{-6}M_\odot$y$^{-1}$.
In each case, when a certain amount of helium is accreted, the triple-alpha
reaction is ignited in the outer part and a shell flash (thermal pulse)
occurs.
The helium burning shell moves inward by repeating 
alternatively shell flashes
and inter-flash compression.
The burning shell reaches the center on a timescale of order $10^6$y
after the first ignition. 
The inward progression does not stop even if the accretion has stopped
after the first shell flash.
At each shell flash a convection zone is formed. When the helium
abundance in the convective layer decreases by $\sim10$\%, the shell
flash diminishes.
When the burning shell reached the center the star has a structure
similar to that of zero-age helium main-sequence star.
Such a star is expected to evolve as a helium star to finally 
become a DB white dwarf. 
We have also discussed the possibility that a low mass double white dwarf
system could be a progenitor of the AM CVn stars.

\acknowledgments
This work has been supported in part by the grant-in-Aid for Scientific
Research (05242102, 06233101, 7223202) and COE research (07CE2002) of the 
Ministry of Education, Science, and Culture in Japan.
KN's research at the Institute for Theoretical Physics has been supported
in part by the National Sceince Foundation in US under Grant No.
PHY94-07194.

\newpage

\begin{figure}
\epsscale{0.8}
\plotone{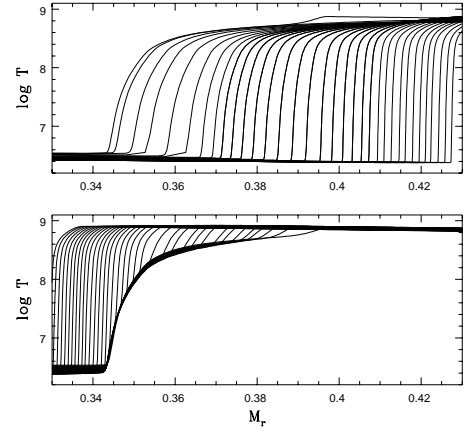}
\caption{Evolution of temperature distribution before (upper panel)
and after (lower panel) the re-activation of flame.}
\label{f7}
\end{figure}

\end{document}